\documentstyle[11pt]{article}
 
\def\pa{\partial}
 
\def\g{\gamma} \def\G{\Gamma}
\def\a{\alpha} 
\def\b{\beta} 
\def\d{\delta} \def\D{\Delta}
\def\e{\epsilon}

\def\l{\lambda} 
\def\m{\mu} 
\def\n{\nu}

\def\s{\sigma}

\def\ul{\underline}
\def\be{\begin{equation}}
\def\ee{\end{equation}}
     
\begin{document}

\begin{flushright}
BRX TH--399
\end{flushright}

\begin{center}
{\large\bf Conformal Anomalies -- Recent Progress}

S. Deser\\
Department of Physics, Brandeis University, Waltham, MA 02254, 
USA\end{center}

\noindent{\bf I. Introduction}

The subject of conformal (or Weyl) anomalies is almost precisely 
20 years old, and has in its lifetime been connected with and 
influenced many important problems in relativity and particle physics, 
from Hawking radiation to conformal field theory and strings, as well 
as mathematics.  The associated literature is correspondingly enormous 
and in this brief review I will concentrate only on the aspects of the 
problem that A. Schwimmer and I \cite{001} as well as others, {\it e.g.}, 
\cite{002,003,004} have been studying recently.  Some of the details 
skipped over here may be found in these references; for some history 
see \cite{005}.

Since the quantum field theoretical background is not familiar to many 
relativists, I will begin with a (very) rapid introduction to anomalies 
as the result of a clash between classical symmetries and the quantum 
requirement of regularization.  This will be illustrated in the 
simplest, 
but as usual, very special case of 2D where everything is unique and 
explicitly presentable in closed form, before going on to explain the 
generalization to four dimensions and higher.  Here we will discuss 
both positive results as well as ``what we know that isn't so," namely 
some widespread misunderstandings about the structure of effective 
gravitational actions, and what we still don't know.  The emphasis 
throughout is on ``classical" aspects that may particularly interest 
relativists.

\noindent{\bf II. Anomalies in General}

Classical matter actions can be endowed with various formal 
invariances.  The classic example here is that of chiral 
anomalies: a ``charged" spinor 
field is invariant both under internal, ``gauge", rotations 
and (up to terms in $m$) under chiral ones involving conjugation 
with $\g_5$.  The corresponding Noether currents are the usual
 $j_\m \sim \bar{\psi} \g_\m 
\psi$ and the chiral current $j_{\m 5} \sim \bar{\psi} \g_5 
\g_\m \psi$.  At the quantum level however, the regularization 
required to define and compute current correlation functions 
involving closed loops (even for the free field!) cannot 
simultaneously preserve both of these invariances
 -- for example, a massive regulator clearly alters the 
chiral current's divergence, while other prescriptions would 
even destroy ``charge" conservation.  All this has no particular 
importance for the free field (since its currents are not 
the sources of anything) but as soon 
as there are even non-dynamical, external, fields present that 
couple to the currents, the consequences become very important 
indeed.  In particular if one considers the closed loop triangle 
diagram represented by the time ordered correlator 
$< T( j_{\m 5} (x) j_\a (y) j_\b (z)) >$, there is a very physical 
effect: each of the $j_a$ is coupled to an 
external $A_\a$, while (the divergence of) $j_{\m 5}$ represents 
a neutral pseudoscalar field (the $\pi^0$).  Thus, the observed 
$\pi^0 \rightarrow 2\g$ 
decay is directly traceable to the -- quantum -- breakdown of 
chiral invariance through the single loop diagram with accompanying 
nonvanishing of the divergence of the above 3-point function 
(of course choosing to break gauge invariance instead would be 
catastrophic!).  Furthermore, although we have here a closed loop, 
and had to regularize to obtain 
a well-defined answer, there are no infinities -- this is a finite 
calculable process.  The anomaly itself is proportional to the 
topological density $F_{\m\n}^{~~*}F^{\m\n}$, {\it i.e.}, 
the chiral current fails to be conserved, by 
$\pa_\m j^{\m 5} \sim \a F^*F$, and there is a corresponding 
effective action expressible in terms of the external fields that 
encodes the ``backreaction" of the quantized matter (fermions) or 
the ``external" pions and photons.

There is an even more obvious arena in which regularization 
through introduction of a mass or of a cutoff or by formally 
continuing away from the physical dimension) destroys an 
invariance, namely that of conformally invariant systems 
involving only dimensionless parameters.  Standard free 
field examples include the Maxwell action (but only at 
$D$=4) or a massless spinor or scalar in any $D$, all of whose 
dilation currents, $D^\m = x^\n T^\m_\n$, are conserved since 
for these systems both
$\pa_\m T^{\m\n} = 0$ and $T^\m_\m = 0$.  [For the scalar 
field in $D > 2$, the usual stress tensor must be suitably 
``improved" in order to become traceless.]  Now one may 
simply follow the same lines as for the chiral anomaly: 
Any regularization introduces a mass or alters the dimension, 
so that the closed loop contributions involving stress-tensor 
vertices lead to vacuum correlation functions involving 
(suitable numbers of) the $T_{\m\n}$ whose quantum invariances 
are diminished -- either conservation or tracelessness is lost.  
Furthermore, one may introduce an external gravitational field 
coupled to the $T_{\m\n}$ vertices in such a way that the anomaly's 
properties may be expressed in purely geometric terms.  The anomaly 
is again \ul{finite} and cutoff-independent and the corresponding 
effective gravitational action that generates it represents the 
back-reaction of matter on the geometry -- hence the connection 
with Hawking radiation.  It is also related to the beta-function 
for the matter system in question (see appendix B), hence the 
special relevance to conformal field and string theory of the 2D 
anomaly.

\noindent{\bf III. Conformal Anomaly in 2D}

Two dimensions are always very special in physics, but there is 
nevertheless a lot to be learned from this simplest context.  
We will see that this is the one case where everything can be 
done explicitly to describe and use the anomaly, and also begin 
to see how higher dimensions will differ in fundamental ways; 
in particular it will become clear what the open problems are,
 and what apparently natural extensions beyond $D$=2 are in 
fact incorrect.

Here and throughout we will use dimensional regularization, 
in which the spacetime dimension is moved by $\e$ from its 
integer value, {\it e.g.}, $D = 2 +\e$, to have entirely 
finite unambiguous correlators before we face the delicate 
question of taking the $\e = 0$ limit.  Near $D$=2, we 
consider in flat space to begin with, the vacuum 2-point correlator, 
\be
K(q)_{\m\n\a\b} = \langle T_{\m\n} (q) \; T_{\a\b} (-q) \rangle \; ,
\ee
where $q$ is the external momentum of this 2-point closed loop 
(and time ordering is understood throughout).  Here $T_{\m\n}$ 
represents the stress tensor operator of one of our massless 
systems, say a scalar, for which $T^\m_\m$ vanishes at any dimension, 
integer or not, by ``improvement".  Specifically, let
\be
T_{\m\n} = ( \phi ,_\m \phi ,_\n - \textstyle{\frac{1}{2}} \:
\eta_{\m\n} \: \phi ,_\a  \phi^{,\a}) 
+ \frac{(D - 2)}{4(1-D)} \:
( \pa_\m \pa_\n - 
\eta_{\m\n} \Box )
\phi^2 \equiv T^0_{\m\n} + \D_{\m\n} \; .
\ee
This is the usual ``minimal" stress tensor $T^0_{\m\n}$ supplemented 
by an identically conserved ``superpotential" term $\D_{\m\n}$ that 
does not affect the Poincar\'{e} generators, and so is allowed.  
On shell $(\Box \phi = 0)$, it is easy to see that $T_{\m\n}$ is 
traceless and conserved, $q^\n \: T^\m_\n (q) = 0$.  Then the 
regulated function $K_{\m\n\a\b}$ must be proportional to 
projectors $P_{\m\n} (q) = (-q_\m q_\n + q^2 \eta_{\m\n})$ on 
each of its indices as well as symmetric under interchange of 
the pairs $(\m\n )$ and $(\a\b )$ and of course traceless in 
each pair.  The unique such form, as also obtained by explicit 
integration over the internal loop momentum is
\be
K_{\m\n\a\b} (q) = f(D)/\e  
\left\{ \left( P_{\m\a} P_{\n\b} + P_{\m\b} P_{\n\a} \right)
- \frac{2}{D-1} \: P_{\m\n} P_{\a\b} \right\}
q^{-2(1+\e )}
\ee
where $f(D)$ is a finite constant depending on the field species.  
Since $P_{\m\a} P_{\a\n} = q^2 \: P_{\m\n}$ and $P^\m_\m = (D-1) 
q^2$, it follow that $K$ obeys all the above requirements.  
Now by power counting (in the loop integration) alone it follows 
{\it a priori} that $K$ must be finite at $D$=2.  We have 
mentioned that finiteness is a hallmark of anomalies, but it 
can be a very subtle one, as we shall see: There must be some 
hidden factor in the numerator to cancel the $\e$ denominators  
at $D$=2.  This is indeed the case for, but only for, the 
$q_\m q_\n q_\a q_\b$ term in (1) which is manifestly proportional 
to $\e$.  For the rest, the mechanism in question is what we have 
called ``0/0" in \cite{001}: exactly at $\e = 0$ the whole 
numerator vanishes identically, as is most easily seen by noting 
that exactly at $D$=2, $P_{\m\n} = \tilde{q}_\m \tilde{q}_\n$, 
$\tilde{q}_\m \equiv \e_{\m\n} q^\n$.  So each term in (3) is 
simply quartic in the $\tilde{q}$'s and their sum vanishes at 
$D$=2. A deeper, and more geometric statement of this comes 
about if we now introduce an external metric and couple each 
$T_{\m\n}$ to this metric in the usual way.  Indeed, it is 
sufficient to use linearized coupling to the deviation $h_{\m\n} 
= g_{\m\n} - \eta_{\m\n}$ from flat space, and then invoke 
covariance to obtain the full answer.  The corresponding functional
\be
W[h_{\m\n}] = \int \int \: h_{\m\n} (x) \langle T_{\m\n}
(x) T_{\a\b} (y) \rangle
h_{\a\b} (y) d^2x \: d^2y
\ee
is of course the (finite) effective gravitational action due
to matter back-reaction (to one-loop order), which incorporates 
the anomaly in the limit $\e \rightarrow 0$.  Using the linearized 
identity (with the sign convention 
$R_{\m\n} \sim + \pa_\a \G^\a_{\m\n})$ 
\be
G^L_{\m\n} = \frac{\Box^{-1}}{2} 
\left( P_{\m\n} P_{\a\b} - P_{\m\a}P_{\n\b} \right) h_{\a\b}
\; , \;\; R^L = P^{\m\n} h_{\m\n} \; , \;\;
P_{\m\n} \equiv (-\eta_{\m\n} \Box + \pa^2_{\m\n})
\ee
we see that
\be
W^L [h] \sim \frac{1}{\e} \: \int \: d^Dx \, d^Dy 
\left[ 4 G^L_{\m\n} \, \Box^{-1} \, G^L_{\m\n} + 
\frac {2\e}{(D-1)} \: R^L \, \Box^{-1} \, R^L \right]
\ee
where we have dropped the (irrelevant) $\Box^\e$ part.  
Now \ul{at} $D$=2, the identity $P_{\m\n} = 
\tilde{q}_\m \tilde{q}_\n$ we found earlier precisely 
implies that $G^L_{\m\n} \equiv 0$, an identity well 
known to be valid to all orders in $h_{\m\n}$, {\it i.e.}, 
for the full Einstein tensor.  Thus, we obtain a meaningful 
prescription for $W^L$ by \ul{defining} the numerator 
to be taken \ul{at} $D$=2, where $G^L_{\m\n} = 0$, which 
leaves the unique finite form
\be
W^L [h] \sim \textstyle{\frac{1}{2}} \int d^2x \, d^2y \,
 R^L (x)
\Box^{-1} (x,y) R^L (y) \; , \;\;\;\;\;\;
\Box [ \Box^{-1} (x,y)] \equiv \d (x-y)\; ,
\ee
for our effective action to lowest order.  Note also the 
single pole structure $\Box^{-1}$, defined of course as 
the flat space scalar propagator (with some choice of 
boundary condition).  This is traceable back to the 
hard-core one-loop Feynman diagram origin of our fancy 
effective action, a fact it will be essential to remember 
also in higher $D$.  Now we can easily improve (7) to a 
fully covariant form, namely
\be
W[g] \sim \int\int d^2x \, d^2y (\sqrt{-g}\, R)(x) 
\langle x | (\sqrt{g} \, \Box )^{-1} |y \rangle (\sqrt{-g} 
\, R) (y)
\ee
in terms of the full curved space propagator indicated.  
This is the celebrated Polyakov action, up to an overall 
(equally celebrated!) coefficient. Let me make two further 
important remarks about $W$.  The first is to remind us 
where the anomalies are: Even though the formal operator 
matter action is Weyl invariant,\footnote{Recall that, in 
a general geometry, flat space conformal invariance is 
promoted to Weyl invariance, where Weyl transformation are 
given by $\d g_{\m\n} = 2\s (x) g_{\m\n} (x)$, $\d \phi (x) 
= \a\s (x)\phi (x)$ and $\a$ indicates the space-time dimension 
of the matter field in question; in 2D a scalar field has 
$\a =0$, while a spinor has $\a = -1/2$, etc.} the resulting 
$W$ is not.  That is, if we vary $g_{\m\n}$ in $W$, using 
the fact that in 2D,
\be
\d (\sqrt{-g} \, \Box ) \equiv  \d (\pa_\m \sqrt{-g} \,
g^{\m\n} \pa_\n ) = 0 \; , \;\;\;\;\;\;
\d (\sqrt{-g} \,R) = -2  \sqrt{-g} \,
\Box \s
\ee
we find that
\be
\d W [g]/\d\s (x) \sim \sqrt{-g} \, R(x) \equiv 
{\cal A} (x)\; ,
\ee
does not vanish.  A corollary is that the anomaly 
${\cal A} (x)$, being the variational derivative of an action, 
must obey the reciprocity relation
\be
\d {\cal A} (x)/\d \s (x^\prime ) = \d {\cal A} 
(x^\prime )/\d \s (x)\; ,
\ee
which it does, since $\d {\cal A}/\d \s^\prime  =
\Box \d^2 (x-x^\prime )= \Box^\prime \d^2 (x-x) = 
\d {\cal A}^\prime / \d\s \; .$

The second observation about this effective action is that 
it contains a single pole; this means in our context an 
excitation of the (hereby induced) gravitational field.  
We can see this by Polyakov's observation that in the 
conformal gauge, $g_{\m\n} = e^{2\phi (x)} \eta_{\m\n}$ 
(always locally reachable in 2D) $W[g]$ reduces to $\int 
d^2x \, \phi \Box \phi$, {\it i.e.}, that its (single) 
Euler--Lagrange equation is $2\Box \phi = - R =0$.  
This 2D characteristic has general validity (see Appendix A).  
We can also notice that the self-interacting $W$ form 
comes from integrating out the scalar field $\phi$ in
\be
W = \int d^2x \, \sqrt{-g} \, R\phi +
 \int d^2x \sqrt{-g} \, \phi \Box \phi \; .
\ee
This is also the Wess--Zumino form of the action: 
suppose we introduce a ``Weyl-compensator" field 
$\phi$ which varies as $\d\phi = \s (x)$.  Then the 
first term gives the desired anomaly $\sqrt{-g} \, R$ 
when we vary $\phi$.  However there is also the extra 
contribution from varying $\sqrt{-g} \, R$ which yields 
$-2 \,\sqrt{-g}\, \Box \phi$; the second term in (12) 
precisely cancels this unwanted piece (recall that $\d 
\sqrt{-g} \, \Box = 0$!).  The form (12) is also obtainable 
by taking the action (8) and subtracting from it its 
Weyl-invariantized version $W[g_{\m\n} e^{-2\phi}]$.  
The expansion in $\phi$ is just (12).

Two final geometric remarks that will be relevant later: 
the first is that there is but one anomaly term possible 
because the integrability condition (11) has only one 
solution with the desired dimensionality -- {\it i.e.}, 
with a local scalar density ${\cal A} (x)$ that is itself 
scale invariant.  The second is that we would have obtained 
the correct $W$, {\it i.e.}, the correct ``numerator" in $W$ 
by using the fact that in 2D any quantity antisymmetric in 
more than 2 indices vanishes, {\it e.g.}, any 
$A^{\a\n}_{[\a\b} A^{\a\b}_{\m\n]} \equiv 0$ where 
${\cal A}_{[\m\n][\a\b]}$ has the algebraic symmetries 
of the Riemann tensor, antisymmetric in each pair and 
symmetric under their interchange; that is in fact the 
useful way \cite{001} to understand higher-dimensional $W$'s.

This pedagogical survey has been intended to illuminate the 
more complicated $D$=4 and higher situations below; 
consequently we skip entirely the  subjects of conformal 
field theory and strings which have the conformal anomaly 
as a base (even any reasonable list of references would 
swamp our text).

\noindent{\bf IV. Four Dimensions}

Let us summarize the lessons from $D$=2: any matter 
system will lead to an effective gravitational action
 through its coupling to external geometry at the 
one-loop level (for free fields, there are of 
course no higher loops!).  If, in particular the 
matter is classically Weyl-invariant, then the 
process of regularization necessarily leads to an 
effective action that is not both diffeo- and 
Weyl-invariant, the anomalous part, ${\cal A} (x) = \d 
W [ge^{2\s (x)}]/\d\s (x) |_{\s = 0}$, being finite 
and local, although the effective action is non-local, 
with the characteristic single denominator $\Box^{-1}$ 
inherited from the loop integral.  In addition it was 
possible to give the full nonlinear form (8) of this 
action, representing the single possible anomaly, the 
Euler density $E_2 \equiv \sqrt{-g} \, R$.  This 
action was furthermore unique; no other 
conformal-invariant functional exists at $D$=2.  
Now we turn to 4D and higher even dimensions 
(anomalies can only occur at even dimensions, as can 
be already understood from the simple fact that no 
local scalar density can exist in odd dimensions that 
is even constant scale invariant).

Let us begin backwards, and ask for a list of 
candidate anomalies, that is scalar densities 
${\cal A} (x)$ that are local, expandable in 
$h_{\m\n}$ (since they can be obtained perturbatively), 
scale invariant and obey the integrability condition 
(11) that permits an effective action (whose 
existence is also perturbatively guaranteed).  
The list here consists of the three independent 
ways to square a curvature, most usefully the combinations
\be
E_4 = \sqrt{-g} \, (R^2_{\m\n\a\b} - 4R^2_{\m\n} + R^2 ) 
\; , \;\;
\sqrt{-g} \, C^2 \; , \;\; \sqrt{-g} \, R^2
\ee
where $C$ is the Weyl tensor and $E_4$ is the 
Gauss--Bonnet topological density that generalizes 
the Euler density $E_2$.  In addition, there is  
what we shall see is a trivial candidate
\be
a = \sqrt{-g} \, \Box R
\ee
and also the Hirzebruch density $R_{\m\n\a\b}^{~~~~*}
R^{\m\n\a\b}$, a parity-odd topological quantity that 
we will not discuss further here except to mention that 
it is itself Weyl invariant (by the cyclic identity 
$R_{[abc]d} = 0$), just like $\sqrt{-g} \, C^2$.  
The Weyl variation of $E_4$ embodies reciprocity since
\be
\d \, E_4 (x)/\d\s(x^\prime ) = G^{\m\n}(x) D_\m D_\n \d 
(x-x^\prime ) =
G^{\m\n}(x^\prime )D_\m^\prime D_\n^\prime \d 
(x-x^\prime ) =
\d E_4 (x^\prime ) / \d \s (x)
\ee
owing to the identical conservation of $G^{\m\n}$.  
In any dimension $D=2n$, $E_{2n}$ being a total divergence, 
will behave like $E_4$ with $G^{\m\n}$ replaced by a higher 
order identically conserved tensor (which, like $G^{\m\n}$, 
vanishes identically in all lower $n$!), so that the 
Euler density is always a legal candidate.  [The proof 
is simple: in components, 
$E_{2n} \sim {\cal E}^{\m_{1} \ldots \m_{2n}} 
{\cal E}^{\n_{1} \ldots \n_{2n}} R_{\m_1\m_2\n_1\n_2} 
\ldots R_{\m_{2n-1}\m_{2n}\n_{2n-1}\n_{2n}}$ 
 with Weyl variation of $R_{\m\n\a\b} 
\sim g_{\m\a} D_\n D_\b \s$ + cyclic, so $\d E_{2n} 
\sim G^{\n\b} D_\n D_\b \, \s$.  It is easy to see 
that $G^{\n\b}$ is the metric variation of $I = \int 
g_{\n\b} G^{\n\b}d^{2n} x$, so it is identically 
conserved, as is also checked directly using the 
Riemann tensor's Bianchi identities.]  This unique 
term we have called type A.  Likewise, at all higher 
dimensions, there will be appropriate generalizations 
$\sqrt{-g} \, C_1 \ldots C_n$ of $\sqrt{-g} \, CC$ at 
$n$=2; these are called type B, and they clearly 
increase in number with dimension since there are 
(at the very least) more independent ways to contract 
indices among the greater number of Weyl tensors.  
Finally, the $\sqrt{-g} \, R^2$ term in (13) is forbidden: 
it fails the integrability test -- obviously there 
is only one identically conserved tensor  linear in 
curvature, namely the Einstein tensor, in 4D.  The 
term $a(x)$ of (14) is integrable, but irrelevant 
because it stems from a purely \ul{local} action such 
as $\int d^4x \sqrt{-g} \, R^2$, a form that is in 
any case needed as a counterterm to the well-known 
``two-point" infinity (rather than the ``three-point" 
nonlocal anomaly).  This pattern persists for all 
$D=2n$: there is either the single, type A, $E_{2n}$ 
term corresponding to the single conserved ``$G^{\m\n}$" 
tensor of rank ($n$-1) (that is the ``Einstein tensor" 
of the action $\int d^{2n}x \sqrt{-g} \, E_{(n-2)} )$ or 
the increasingly large type B set of Weyl invariants 
$\sqrt{-g} \, C^n$ just discussed, in addition to local 
anomalies. [There is very nice agreement, incidentally, 
between the 
present analysis and cohomology arguments such as those 
of \cite{006} and references therein.]
The specific coefficients of the various anomaly terms 
have been tabulated for all massless free fields (in $D$=2, 
4 at least), in terms of their spin content, but there 
is more, for interacting systems, that involves their 
$\b$-functions.  Here I only have space to sketch the 
effective action problem and some attractive, but alas 
invalid, closed form solutions of it.

Let us first dispose of the action problem for type B: 
what $W[g]$ gives $\sqrt{-g} \, C^2$ upon Weyl variation?  
Clearly, since $\sqrt{-g} \, C^2$ is itself inert, we 
want something of the form $W_B \sim \int d^4x \sqrt{-g} 
\, C^2 X$ where $X$ is a scalar that varies as $\s (x)$, 
to linear order, say.  The only such diffeo-invariant 
candidates ({\it i.e.}, scalars) are (to linearized order 
in $h_{\m\n}$) $R/\Box$ and ln $\Box$; they have 
profoundly different origins in terms of scale dependence, 
and the (only) correct choice is \cite{001} ln $\Box$.  
Indeed, this was the first nonlocal anomaly to be discovered 
in $D$=4, and the correct $W_B$ was already given there; 
nevertheless the wrong choice has often cropped up since.  
The definition of the nonlocal ln $\Box$ (more correctly 
ln $\Box /\m^2$, where $\m$ is a regularization scale) 
is straightforward and, to the operative cubic order in 
$h_{\m\n}$, its location in the integral may simply be 
taken  ``between" the two Weyl tensors.  The closed form 
extension of this action is not known explicitly, but must 
exist.

In type A, we need a $W_A [g]$ with a single pole; to 
lowest (cubic) order in $h_{\m\n}$, it has the somewhat 
inelegant form \cite{001}
\be
W_{3,d=4} = \int d^4x \sqrt{-g} \, \Box^{-1} \left[ 
\frac{1}{2}  \, R^2_{\m\n\a\b} R + 10 R_{\m\n}R^{\n\a}
R^\m_\a - 13 R^2_{\m\n}\, R + \frac{41}{18} \, R^3 + 
6R_{\m\n\a\b} R^{\m\a}R^{\n\b} \right]
\ee
which, however, derives in a direct way from the ``0/0" 
ideas of $D$=2 by using the $\e \rightarrow 0$ limit of 
the cubic form $C^{\m\n}_{[\m\n} C^{\a\b}_{\a\b} 
C^{\l\s}_{\l\s]}$ together with a ``floating" $\Box^{-1}$ 
that for present purposes can be between any two of the 
factors.  It is essential to a correct prescription that 
it be consistent with the known field-theoretic rules for 
anomalies, as well as with, of course, the Ward identities; 
this is fulfilled by (16). [Indeed, to get $E_4$ from 
varying (16) required frequent use of this apparatus!] 
To date, we have been unable to find a closed form, 
however.  There does exist a very elegant closed form 
expression whose Weyl variation simply yields $E_4$; 
the only problem is that is is wrong, {\it i.e.}, it 
cannot arise from a loop integral.  The form in question 
rests on a simple analogy with the Polyakov form (8) in 
2D; there, $\d_\s \, E_2 = (\sqrt{-g} \, \Box ) \s$ while 
$\d_\s (\sqrt{-g} \, \Box ) = 0$, which immediately 
justifies (8).  Can this be promoted to $D=4$?  First, 
it is clear that the analog of $\sqrt{-g} \, \Box$ 
must be something like $\sqrt{-g} \, \Box^2$ in order 
to be even constant scale invariant (let alone Weyl 
invariant).  Indeed, the correct operator is 
\be
\D = \Box^2 + 2D_\m (R^{\m\n} - 
\textstyle{\frac{1}{3}} \, g^{\m\n} R) D_\n
\ee
a fairly simple (self-adjoint) generalization. 
[At 6D and beyond such a $\D \sim \Box^n  +\ldots$ 
also exists but is no longer unique \cite{002}]. 
Likewise, while $E_4$ does not quite vary correctly, 
the quantity
\be
\bar{E}_4 \equiv (E_4 - \textstyle{\frac{2}{3}} \,
 \Box R ) \; , \;\;\;
\d \bar{E}_4 = \sqrt{-g} \, \D \s
\ee
does, {\it i.e.},
\be
\d \left( \int \bar{E}_4 \D (\sqrt{-g} \, \D )^{-1} 
\bar{E}_4 \right) / \d \s (x) = \bar{E}_4 (x) \; .
\ee
This differs from the desired $E_4$ by a local 
anomaly $\frac{2}{3} \, \Box R$, which means that 
the  action is rather the one varied below:
\be
\d \left( \frac{1}{2} \int \bar{E}_4 (\sqrt{-g} \, 
\D )^{-1} \bar{E}_4 - \frac{1}{18} \, \int R^2 \right) 
= E_4 \; .
\ee
The trouble with this form, however, is that it has 
a double pole, already to order $h^3$, and hence is 
not viable.  Likewise, the form
$\int \bar{E}_4 (\sqrt{-g} \, \D )^{-1} \: \sqrt{-g} \, C^2$, 
whose variation gives $\sqrt{-g} \, C^2$, has the wrong 
scale dependence.  In fact its lowest (cubic) part is just 
$\sim \int C^2 \, R/\Box$ which is incorrect.  What is also 
interesting is that the ``bad" type A form (20) is equivalent 
to the Wess--Zumino (WZ) expression that yields $E_4$, so that 
here too the 2D reasoning fails.  Rather than give the 
mechanism behind the general WZ construction that ``mechanically" 
yields the desired action, the result is sufficiently simple 
that we can reach it iteratively.  We start by introducing 
the Weyl compensator field $X,\;\d X = \s (x)$ and with the 
obvious zeroth ansatz
\be
W_0 = \int d^4x \: E_4 X \; .
\ee
We must, however, compensate for the fact that the 
Weyl variation of $E_4$ gives the unwanted contribution 
$G^{\m\n} D_\m D_\n X$ to ${\cal A} (x)$ by adding
\be
W_1 = \frac{1}{2} \: \int \int d^4x \: G^{\m\n} D_\m X \:
D_\n X \; .
\ee
Now, however, we get an unwanted contribution from $\d 
G^{\m\n} \sim (D^\m D^\n - g^{\m\n} \Box )(D_\m X D_\n X )$
that requires a cubic term $W_2 \sim (D_\m X)^2 \Box X$, 
and finally a quartic term
$W_3 \sim \int d^4x \sqrt{g} \, (D_\m X)^2 (D_\n X)^2$ is
 needed to cancel the contribution from $\d (\Box )$ 
in $W_2$.  The full closed form WZ action is then
the appropriate sum,
\be
W_{\rm WZ} = \int\int d^4x \left\{ E_4  X + a G_{\m\n} 
D_\m X \,
D_\n X + b \Box X(D_\m X)^2 + c
[(D_\m X )^2 ]^2 \right\} \; .
\ee
Unlike its 2D counterpart, however, this form is 
neither Gaussian, nor does it even have a kinetic 
term $\sim \int X \Box X$ at all, so we cannot go 
from it to a closed form, and setting $X \sim R/\Box$ 
as a lowest approximation introduces unacceptable 
$\Box^{-2}$ terms.  This is not surprising, because 
one can show that this $W_{\rm WZ}$ (23) is closely 
related to the ``Polyakov" expression (20) and its $W$.  
Indeed, one can show that (23) is just
\be
W_{\rm WZ} = W [g] - W [ge^{-2X} ] \; .
\ee
The last term on the right side being manifestly 
Weyl invariant, the two clearly yield the same anomaly.

At this point, then, we have two different $D$=4 
actions, to leading (cubic) order about flat space 
for both type A and B anomalies, but only one correctly 
reproduces the underlying loop physics.  In 2D, where 
there was only type A, this action was furthermore 
unique; no $\D W$ can be constructed that is Weyl 
invariant.  Thus, knowledge of the anomaly determined the 
whole effective gravitational action there (see Appendix A).  
[Of course, a less impressive way to say this is that since 
the general 2D metric is conformally flat, only the $\d W /
\d \s (x)$ is relevant anyway!]  Is there similar uniqueness 
in $D \geq 4$, {\it i.e.}, do we expect that knowledge of 
the anomaly also determines the effective action here?  
From the above parenthetic remark, we should expect a 
negative answer.  Indeed, let us show how to construct at least 
type A-like (with $\Box^{-1}$ behavior) 
$\D W$'s that are Weyl invariant to the same, lowest, 
order as that of our $W$ of (16) itself.  The idea is 
very simple.  Consider in 4D the local cubic Weyl invariants, 
which are in fact the known type B anomalies in 6D.  Although 
there are 2 such invariants in 6D (namely the apparently 
different ways of tracing the product of 3 Weyl tensors), 
they are equivalent in 4D owing to the identity 
$C^{\m\n}_{[\m\n} \, C^{\a\b}_{\a\b} \, C^{\l\s}_{\l\a ]} 
\equiv 0$ 
here (6 indices are antisymmetrized).  Thus the action 
$\D W = \int d^4x \,{\rm tr} C^3/\Box$ is clearly Weyl 
invariant to lowest order 
in $h_{\m\n}$.  [The reason one cannot use the same idea 
in 2D, with $C^2$ as the invariant is of course that the 
Weyl tensor vanishes identically here.]  Whether these 
leading order Weyl invariants really persist to all 
orders is not immediately clear, though there is no 
reason to doubt it (the overall scale invariance is 
formally preserved by $\int d^4x \sqrt{-g} \, C^3/\Box$ 
for example).\footnote{Another possible set of ambiguities 
derive from a different set of 6D integral Weyl invariants 
that begins with terms like $C\Box \!C$ plus cubic 
curvatures.  There are two such,  given in \cite{002}.  
[One of them 
is found in (25c) of \cite{001}, but two corrections 
must be made there: the relative sign of the $C\Box \! 
C$ and of the remaining terms is wrong and only the 
integral, not the density itself is Weyl invariant.]  
It is conceivable that the 4D integral of these 
quantities divided by the $\Box$ operator may also 
have an invariance to cubic order; the quadratic 
part from $C\Box \! C$ for example reduces to the 
(irrelevant)local invariant~$\int d^4x \,\sqrt{-g} 
C^2$.}   On the other hand, there seems to be no 
ambiguity in the type B actions, involving the ln $\Box$ 
factors.

That there is room for ambiguity does not of course 
mean that it is always present; indeed a very recent 
interesting paper \cite{007} on 4D conformal systems 
(of a very special type) derived the gravitational 
action uniquely from the anomaly. However, this 
uniqueness is probably related to the higher symmetry 
(K\"{a}hler structure) of the 4-manifold there. 
In any case what is really important is whether 
the coefficients of the type A effective action 
can be exploited as in 2D CFT to relate different 
conformal systems.

I am grateful to my coauthor, A. Schwimmer with whom  
this research was carried out.  I also thank R. 
Palais for a useful conversation on Appendix A.  
This work was supported by the National Science 
Foundation, grant \#PHY-9315811, and initially 
also by a US--Israel BSF grant.

\newpage

\begin{center}
{\large\bf Appendix A}

{\bf Varying 2D gravitational actions}
\end{center}

Strictly speaking, one cannot first fix a 
gauge in an action, and then deduce the 
field equations by varying the remaining 
field components in a gauge theory; one 
would then in general miss the constraints, 
such as the Gauss law (fixing $A_0 = 0$) 
or the Hamiltonian constraints in Einstein 
theory (fixing $g_{0\m} = \eta_{0\m}$).  
However, it is intuitively clear that since 
there is  only one independent metric 
component in 2D, it must be an exception
and that the three field equations 
$\d I/\d g_{\m\n} = 0$ reduce to only one ``real" 
one and two ``Bianchi identities".  For orientation, 
consider first the Polyakov action in  linearized 
approximation.  In terms of the variables 
in a 1+1 decomposition, 
$h \equiv h_{11}, \; N = h_{00}, \; L \equiv h_{01}$, 
the linearized curvature is
$$
R^L \equiv \pa^2_{\a\b}h^{\a\b} - \Box h^\a_\a = 
(h^{\prime\prime} - 2\dot{L}^\prime + \ddot{N}) - 
(h-N)^{\prime\prime} + (\ddot{h}-\ddot{N})
= \ddot{h} + N^{\prime\prime} - 2\dot{L}^\prime
$$
and consequently varying either the constraint 
variables $(N,L)$ or the ``dynamical" $h$ will 
give the same $\Box^{-1} R^L = 0$ equation, 
which implies $R^L =0$.  The content of this 
equation is of course most obvious in conformal 
gauge, $h_{\m\n} = \phi \:\eta_{\m\n}$, where 
$R^L (\phi \: \eta_{\m\n}) \equiv - \Box \phi$.  
Because there is only one independent equation 
here, it would naturally also have been found 
by immediately fixing the gauge in 
$\int\int R^L \Box^{-1} R^L \rightarrow \int d^2x \, 
\phi \Box \phi$ and varying to find $\Box \phi = 
-R = 0$.  In covariant form, varying (8) gives 
$(\pa_\m\pa_\n - \eta_{\m\n} \Box )(R/\Box ) = 0$, 
whose trace part is indeed $\sim \Box (R/ \Box ) = 0$.  
The remaining two components are automatically satisfied 
by $R=0$; their separate content is that $\pa_t \pa_x 
(R /\Box ) = 0 = (\pa^2_t + \pa^2_x) (R/\Box )$, which 
is only formally (a bit) stronger than $\Box (R/\Box ) 
= 0$. 
In the nonlinear case, this is less evident because 
the $(\sqrt{-g} \, \Box )^{-1}$ factor actually 
depends on different combinations of the metric, 
but the result nevertheless is valid: 
To justify it, let us vary  the full  Polyakov 
action (8) under all $\d g_{\m\n}$.  The variation 
of the $(\sqrt{g} \, \Box )^{-1}$ factor yields a 
term $\D_{\m\n}$ that is identically traceless, 
since $\sqrt{g} \, \Box$ depends only on the 
unimodular combination, $\sqrt{-g} \, g^{\m\n}$.  
More specifically, $\D_{\m\n}$ has the form of 
the usual scalar field's stress tensor with 
$\phi \sim R/\Box$.  The variation of $\sqrt{-g} \, 
R$ yields the form 
$(D_\m D_\n - g_{\m\n} \Box )(R/\Box )$, whose 
trace is just $R$ itself, as expected.  Hence the 
trace of the full field equation already implies
$R$=0, thereby automatically fulfilling the other 
two (traceless) components of the equations, modulo 
the formal point just made for the linear case.
The above result, that conformal variation gives 
all the information is less obvious for other 2D 
actions, even for the local  $I = \int d^2x \sqrt{-g} 
\, R^2$.  Its Euler equations, using $R_{\m\n} = 
\frac{1}{2} \, g_{\m\n} R$, are ${\cal G}^{\m\n}
\equiv (-D_\m D_\n + g_{\m\n} \Box ) R + 
\frac{1}{2} g_{\m\n} R^2 = 0$ which obey the Bianchi 
identities $D_\n {\cal G}^{\m\n} \equiv 0$.  
The trace equation is $\Box R + R^2 = 0$, 
which seems to be weaker than the tensorial 
${\cal G}_{\m\n} = 0$; but we know from 
our linearized discussion that they really 
are not. In conformal gauge, of course, $I = 
4 \int d^2x e^{-2\phi} (\Box \phi )^2$, 
leading to the single equation for $\phi$ that 
represents the above trace.
\newpage

\begin{center}
{\large\bf Appendix B}

{\bf Type B and $\b$ functions}
\end{center}

As an example of the relation between the 
type B anomaly and $\b$-functions, which also 
brings out the role of a scale in type B, 
and how the ``invariance clash" is seen at 
a simple diagrammatic level, we take 4D 
self-interacting $\phi^4$ theory, which is of 
course classically scale invariant.  The relevant 
(Fourier transformed) correlators
$$
L_{\m\n} (q; k, p) \equiv \langle 
T(T_{\m\n}(q) \phi^2 (k_1) \phi^2 (k_2)) 
\rangle \;\;\;\;\;\;\; K(k) = \langle  
T(\phi^2 (k)\phi^2 (-k)) \rangle
$$
are not purely among stress tensors, but 
instead represent the triangle with one 
graviton and two $\phi^4$ corners, and the pure 
scalar 2-point loop respectively.  [There is 
also a contact term where a graviton emerges 
from one of the 2-point loop's ends, but that 
can be redefined into $L_{\m\n}$ by appropriate 
subtraction.]  There are then two separate Ward 
identities representing (linearized) Weyl and 
coordinate invariance, and they cannot both be 
maintained -- one signal is that $K$ is 
logarithmically divergent and hence requires 
introduction of a scale: $K(k^2) = \ln \: k^2/\m^2$.  
Decomposing $L_{\m\n}$ into invariant amplitudes 
after Fourier transforming and expanding 
in tensorial combinations of the two external 
momenta, one finds three relations among the 
four independent amplitudes, and that the UV 
divergences embodied by the cutoff in $K$ cancel, 
because only $K(k^2_2) - K(k^2_1)$ enters.  The 
invariance clash is best seen by going to a special 
point in momentum space where $(k_1 + k_2)^2 = 0, \; 
k^2_1 = k^2_2$.  There, one discovers that the same 
structure function is simultaneously constrained by 
the respective Ward identities both to vanish and to 
be proportional to $k^2 d \, K/d k^2$.  Explicit 
diagrammatic calculations confirm that $K$ indeed  
has a $\d$-function discontinuity.  Choosing to 
preserve conservation, then, has resulted in the 
conclusion that $T^\m_\m$ is proportional to $\l \phi^4$, 
{\it i.e.}, to the beta function of the theory.  
Note that this whole calculation of the beta function 
has been entirely in the infrared domain and does not 
involve UV properties.  Similar results have also been 
found very recently in the second paper of \cite{003}.


\begin{thebibliography}{99}
\bibitem{001} S. Deser and A. Schwimmer, Phys. 
Lett. {\bf B309} (1993) 279.
\bibitem{002} D.R. Karakhanyan, R.P. Manvelyan, 
and R.L. Mkrtchyan, Yerevan preprint (1996).
\bibitem{003} H. Osborn and A. Petkou, Ann. 
Phys. {\bf 231} (1994) 311; J. Erdmenger and 
H. Osborn, hep-th 9605009 preprint.
\bibitem{004} L. O'Raifeartaigh, I. Sachs, 
and C. Wiesendanger, hep-th 9607110 preprint.
\bibitem{005}
M.J. Duff, Class. Quant. Grav. {\bf 11} (1994) 
1387.
\bibitem{006}
L. Bonora, P. Pasti and M. Bregola, 
Class. Quant. Grav. {\bf 3} (1986) 635.
\bibitem{007} A. Losev, G. Moore, N. Nekrasov, 
and S. Shatashvili, hep-th 9606082 preprint.

\end{thebibliography}
\end{document}